\documentclass[acmsmall,nonacm]{acmart}
\AtBeginDocument{%
  }

\setcopyright{acmlicensed}
\copyrightyear{2018}
\acmYear{2018}
\acmDOI{XXXXXXX.XXXXXXX}

\acmJournal{JACM}
\acmVolume{37}
\acmNumber{4}
\acmMonth{8}

\usepackage{minted}
\usepackage{listings, listings-module}
\usepackage[dvipsnames]{xcolor}
\usepackage{pifont}
\usepackage{caption}
\usepackage{multirow}

\usepackage[labelformat=simple]{subcaption}

\usepackage{tabularx}
\usepackage{ltablex}
\usepackage{enumitem}

\usepackage{tikz}
\usetikzlibrary{shapes,calc}

\usepackage{tcolorbox}

\newcommand{\cmark}{\textcolor{ForestGreen}{\ding{51}}}
\newcommand{\xmark}{\textcolor{red}{\ding{55}}}
\newcommand{\qmark}{\textcolor{Cerulean}{\textbf{?}}}

\newcommand{\system}{Astrogator}

\makeatletter
\newcommand{\curcord}{\the\tikz@lastxsaved,\the\tikz@lastysaved}
\makeatother

\newcommand{\den}[1]{\left\llbracket #1 \right\rrbracket}

\usepackage{soul}
\setul{1pt}{.4pt}

\newcommand{\vrb}[1]{\textcolor{Red}{\textbf{#1}}}
\newcommand{\desc}[1]{\textcolor{NavyBlue}{\text{\ul{#1}}}}
\newcommand{\sep}[1]{\textcolor{OliveGreen}{\textit{#1}}}
\newcommand{\args}[1]{\textcolor{Mulberry}{\textit{\ul{#1}}}}

\newcommand{\figsize}{\footnotesize} 

\begin{document}

\title{Towards Formal Verification of LLM-Generated Code from Natural Language Prompts}

\author{Aaron Councilman}
\email{aaronjc4@illinois.edu}
\affiliation{%
  \institution{University of Illinois at Urbana-Champaign}
  \city{Urbana}
  \state{Illinois}
  \country{USA}
}

\author{David Jiahao Fu}
\email{jiahaof4@illinois.edu}
\affiliation{%
  \institution{University of Illinois at Urbana-Champaign}
  \city{Urbana}
  \state{Illinois}
  \country{USA}
}

\author{Aryan Gupta}
\email{aryan10@illinois.edu}
\affiliation{%
  \institution{University of Illinois at Urbana-Champaign}
  \city{Urbana}
  \state{Illinois}
  \country{USA}
}

\author{Chengxiao Wang}
\email{cw124@illinois.edu}
\affiliation{%
  \institution{University of Illinois at Urbana-Champaign}
  \city{Urbana}
  \state{Illinois}
  \country{USA}
}

\author{David Grove}
\email{groved@us.ibm.com}
\affiliation{%
  \institution{IBM Research}
  \city{Yorktown Heights}
  \state{New York}
  \country{USA}
}

\author{Yu-Xiong Wang}
\email{yxw@illinois.edu}
\affiliation{%
  \institution{University of Illinois at Urbana-Champaign}
  \city{Urbana}
  \state{Illinois}
  \country{USA}
}

\author{Vikram Adve}
\email{vadve@illinois.edu}
\affiliation{%
  \institution{University of Illinois at Urbana-Champaign}
  \city{Urbana}
  \state{Illinois}
  \country{USA}
}

\renewcommand{\shortauthors}{Councilman et al.}

\begin{abstract}
In the past few years LLMs have emerged as a tool that can aid programmers by taking natural language descriptions and generating code based on it.
However, the reliability of LLM code generation and current validation techniques for it are far from strong enough to be used for mission-critical or safety-critical applications.
In this work we explore ways to offer formal guarantees of correctness to LLM generated code; such guarantees could improve the quality of general AI Code Assistants and support their use for critical applications.
To address this challenge we propose to incorporate a Formal Query Language that can represent a user's intent in a formally defined but natural language-like manner that a user can confirm matches their intent.
We then have a formal specification of the user intent which we can use to verify that LLM-generated code matches the user's intent.
We implement these ideas in our system, \system{}, for the Ansible programming language, widely used for system administration, including for critical systems.
The system includes an intuitive formal query language, a calculus for representing the behavior of Ansible programs, and a symbolic interpreter and a unification algorithm which together are used for the verification.
A key innovation in \system{} is the use of a Knowledge Base to capture system-specific implementation dependencies that greatly reduce the need for system knowledge in expressing formal queries.
On a benchmark suite of 21 code-generation tasks, our verifier is able to verify correct code in 83\% of cases and identify incorrect code in 92\%.
\end{abstract}




\maketitle

\section{Introduction}
In recent years, Large Language Models (LLMs) have proven to be surprisingly successful at generating code from natural language prompts~\cite{code-llama.arxiv23, deepseek-coder.arxiv24, starcoderv2.arxiv24}.
The apparent success of LLMs at this task has led some to speculate that prompting LLMs to generate code will fully replace traditional programming, but programming in natural language has many challenges~\cite{foolishness.dijkstra79} and using LLMs for this purpose has even more~\cite{premature.cacm23}.
For example, LLMs are known to suffer from hallucinations in code generation, including generating code that uses non-existent APIs and libraries~\cite{codegen-hallucinations.issta25} and LLMs have also been shown to struggle with many practical programming tasks that rely on complex functions~\cite{zhuo_bigcodebench_2025, nguyen_codemmlu_2025} and involve reasoning about large codebases~\cite{swebench.iclr24}.
Even when the generated code is valid, studies on AI Code Assistants suggest that users have a hard time checking LLM generated code~\cite{usabilityCodeAssistant.icse24}, that bugs in LLM generated code are often subtle~\cite{dou_whats_2024}, and the use of AI Coding Assistants can be associated with less secure code~\cite{insecure.ccs23}.
For mission-critical and safety-critical (together ``critical'') software, such as software in network stacks, distributed systems protocols, financial software, and embedded systems controllers for critical systems, the use of LLM code generation can pose serious risks as errors can have disastrous impacts.
The literature on AI Code Assistants suggests they offer potential benefits to programmers~\cite{productivityCodeAssistant.MAPS22,usabilityCodeAssistant.icse24,groundedCopilot.oopsla23}
and so the appeal of using them for critical code is apparent, especially as critical software is often difficult and complex.
However, because of the significant risks that errors can pose, any code in these systems needs to be correct and significant efforts in formal verification have been focused on such systems~\cite{compcert.popl06,sel4.sosp09,smartcontracts.pmc20}.
As such, it is essential to provide formal guarantees about any LLM generated code for such systems.

While using tests has been explored as a way to check LLM-generated programs~\cite{codex.arxiv21,codet.iclr23} this approach is not sufficient in this context because tests can only guarantee the correctness of the program on the finite set of tested inputs, meaning test cases can \emph{never} completely guarantee correctness.
In critical systems, a guarantee of correctness only a subset of possible inputs is not sufficient because it means failure may be possible.
Additionally, the assurance of correctness that tests can offer is highly dependent on their quality~\cite{testquality.cme21,testmistakes.stvv11}, including factors like how well they cover the space of possible inputs (including invalid or unexpected inputs) and coverage of the code; because of the many factors that must be considered, developing good tests is difficult~\cite{testing.art3rd}.
These challenges are exacerbated in settings like system configuration tasks, which can be critical software if used to configure such systems, because identifying how to test something can be very difficult; as an example, how do we test that the Python package \textsf{numpy} is installed?
It might seem that running the Python command \texttt{import numpy} is sufficient but this will succeed if there is simply a \texttt{numpy.py} file in the current directory.
For these reasons, tests are not sufficient and not a replacement for formal guarantees and formal verification of critical software.

\emph{Our broad, long-term, goal then is to provide a formal guarantee that code generated by an LLM from a natural language prompt has the user's desired functionality}.
This is a very hard problem that poses several significant challenges.
Perhaps the most fundamental is that there is no formal specification: the natural language prompt is inherently informal and may be ambiguous, vague, or incomplete.
This means traditional approaches that provide formal guarantees, such as program synthesis and verification, are not immediately applicable:
without a formal specification, what property would one prove?
Second, even if a formal specification were available, proving the correctness of LLM-generated code with respect to that specification poses additional challenges.
One approach to \textit{prove} correctness against a formal specification is to use symbolic execution.
This requires the formal semantics of both the target and formal specification languages.
It also requires defining a symbolic execution engine and translating the semantics of both languages to the input language of the execution engine.
Finally, it requires defining a verification procedure that can prove equivalence between the two executions, a task that is often difficult to make scalable.

In this work, we propose a general solution to the the first problem: of how to obtain a reliable formal semantics for the user's intent.
We also address the other challenges in the somewhat limited setting of a class of Domain Specific Languages.
These other challenges build on a long history of work in formal verification and we leave it to future work to apply these techniques to broader classes of languages.

To address the first problem, we propose to add a high-level \textbf{formal query language} which is designed to be close to the syntax of natural language but is precisely and formally defined.
We aim to design this formal query language to be at a similar level of abstraction to that of the user's natural language description --- using similar concepts and abstractions --- and to make it readable and intuitive, so that a user can easily understand the formal query and ensure it reflects their intent.
This \emph{user-validated formal query} then gives us a \emph{reliable formal specification of the user's intent} we can use to verify the LLM-generated programs.
While verification remains a very difficult problem, this work offers a bridge between the user's intent and the realm of formal verification.

As a first step on the rest of the challenges, we focus on Domain-Specific Languages (DSLs), which LLMs often struggle to generate~\cite{Gu_2025}.
We believe that providing formal guarantees of correctness may be easier for DSLs because we can use specialized domain knowledge not available in general-purpose programming languages and, since many DSLs are not Turing Complete, some techniques from program verification may be more tractable.
In this work, we focus on the Ansible programming language, which is used for IT automation, where correctness can be critical when used to configure critical systems.
Ansible has a relatively large user-base and open source community and was acquired by Red Hat in 2015.
We also demonstrate, through examples, that our techniques to offer guarantees for LLM generated code can be used for other languages, specifically Bash and a microcontroller programming language (see Section~\ref{sec:generality}) but we leave a full development of verification systems for these languages to future work.

Our key contributions in this work are as follows:
\begin{enumerate}
    \item Formalize our notion of programming using natural language and guaranteeing the code's correctness (Section~\ref{sec:problem-statement}).
    \item Propose the idea of an (artificially introduced) formal query language that captures the user's intent in precise but high-level terms that can be reviewed by the user (Section~\ref{sec:fql-theory}).
    \item Present \system{}, an implementation of such a formal query language and verifier for the Ansible programming language (Section~\ref{sec:ansible-impl}).
    \item Evaluate \system{} on a benchmark suite of 21 LLM code-generation tasks for Ansible programs from natural language descriptions, showing that it is able to formally prove correctness of generated code in 83\% of cases, and to precisely identify incorrect LLM generated code in 92\% of cases. We also detail the failures of \system{} and show that they are due to minor limitations in our implementation and testing set-up, and not due to limitations of the overall approach (Section~\ref{sec:evaluation}).
\end{enumerate}

The other sections of this paper discuss background on Ansible and errors in LLM-generated Ansible code (Section~\ref{sec:background}), the generalizability of our approach (Section~\ref{sec:discussion}), comparisons with related work (Section~\ref{sec:related-work}), and our conclusions (Section~\ref{sec:conclusion}).

\section{Background}\label{sec:background}
\subsection{Ansible}\label{sec:background-ansible}
Ansible is an open-source system used to automate many IT tasks involved in the management of remote systems (such as servers or virtual machines) including provisioning, system configuration, and application deployment and orchestration~\cite{ansible.redhat,ansible.starting}.
Ansible runs on one system, known as the \emph{controller}, connects to the remote systems being managed via SSH, and sends commands from the controller to the remote systems and information and responses back.
This simple architecture enables Ansible to be used for a wide variety of remote systems.

\begin{figure}
    \centering
    \begin{minipage}{.49\textwidth}
\begin{minted}[
linenos=true, fontsize=\scriptsize, frame=lines, escapeinside=!!,
]{yaml}
---
- name: Install and start apache on servers !$\label{line:sample-play}$!
  hosts: servers !$\label{line:sample-hosts}$!
  become_user: root !$\label{line:sample-become}$!
  become: true !$\label{line:sample-play-c}$!
  tasks: !$\label{line:sample-tasks}$!
    - name: Install apache server !$\label{line:sample-task1}$!
      ansible.builtin.package:
        name: "{{ 'apache2' if ansible_os_family
                    == 'Debian' else 'httpd' }}"
        state: present !$\label{line:sample-task1-end}$!
    - name: Start apache service !$\label{line:sample-task2}$!
      ansible.builtin.service:
        name: "{{ 'apache2' if ansible_os_family
                    == 'Debian' else 'httpd' }}"
        state: started !$\label{line:sample-task2-end}$!
\end{minted}
        \caption{An example Ansible playbook composed of a single play which configures a set servers by installing and starting Apache server.}
        \label{fig:ansible-sample}
    \end{minipage}%
    \hfill%
    \begin{minipage}{.49\textwidth}
    \scalebox{.6}{
    \begin{tikzpicture}[component/.style={rectangle, minimum width=2.5cm, minimum height=1cm}]
        \node[] (user) {};
        \node[circle,draw=black,minimum width=.9cm,fill=black, above of=user, node distance=0.5cm] (head) {};
        \node[semicircle, draw=black, below of=user, node distance=0.5cm, minimum width=1.5cm, fill=black] (body) {};
        \node[below of=user, node distance=1cm] (username) {User};

        \node (userinput) at ($(username.north) + (0, 0.3cm)$) {};

        \node[below of=user, node distance=3cm] (systemloc) {};

        \node[component, draw=green, fill=green!10, left of=systemloc, node distance=4cm] (formalizer) {Formalizer};
        \node[component, draw=green, fill=green!10, right of=formalizer, node distance=4cm] (querycheck) {Query Check};

        \node[component, draw=red, fill=red!10, below of=querycheck, node distance=2cm] (codegen) {Code Generator};

        \node[component, draw=blue, fill=blue!10, right of=codegen, node distance=4cm] (verifier) {Verifier};
        \node[component, draw=blue, fill=blue!10, right of=querycheck, node distance=3cm, align=center] (assumptioncheck) {Assumptions\\Check};

        \draw[-latex,thick,color=ForestGreen] ($(userinput) - (1cm, 0)$) -- (userinput -| formalizer) -- (formalizer) node[midway,right,yshift=-.2cm]{\small NL Description};
        \path[-latex,thick,color=blue] ($(formalizer.east) + (0, 0.1cm)$) edge node[above,align=center]{\small Formal\\Query} ($(querycheck.west) + (0, 0.1cm)$);
        \path[-latex,dashed,color=gray] (username.south -| querycheck) edge (querycheck);
        \path[-latex, color=gray] ($(querycheck.west) - (0, 0.1cm)$) edge node[below,at start,xshift=-0.2cm]{\small\xmark} ($(formalizer.east) - (0, 0.1cm)$);
        \draw[-latex,thick,color=ForestGreen] ($(formalizer.north) + (0, 0.4cm)$) -- ($(formalizer.north west) + (-0.4cm,0.4cm)$) -- (\curcord |- codegen) -- (codegen);
        \path[-latex,thick,color=blue] (querycheck) edge node[left,at start,yshift=-0.2cm]{\small\cmark} (codegen);
        \path[-latex,thick,color=red] ($(codegen.east) + (0, 0.1cm)$) edge node[above]{\small Code} ($(verifier.west) + (0, 0.1cm)$);
        \draw[-latex,thick,color=blue] ($(codegen.north) + (0, 0.4cm)$) -- ($(verifier.north west) + (0.2cm, 0.4cm)$) -- (\curcord |- verifier.north);
        \path[-latex,color=gray] ($(verifier.west) - (0, 0.1cm)$) edge node[below,at start,xshift=-0.2cm]{\small\xmark} ($(codegen.east) - (0, 0.1cm)$);
        \path[-latex,thick,color=Cerulean] ($(verifier.north) - (0.2cm, 0)$) edge node[left,at start,yshift=0.2cm]{\small\qmark} (\curcord |- assumptioncheck.south);
        \draw[-latex,dashed,color=gray] ($(userinput) + (1cm, 0)$) -- (userinput -| assumptioncheck) -- (assumptioncheck);
        \path[-latex,thick,color=red] ($(verifier.north) + (0.6cm, 0)$) edge node[right,at start,yshift=0.2cm]{\small\cmark} (\curcord |- userinput);
        \path[-latex,thick,color=red] ($(assumptioncheck.north east) - (0.1cm, 0)$) edge node[left,at start,yshift=0.2cm]{\small\cmark} (\curcord |- userinput);
        \path[color=gray] ($(assumptioncheck.south) - (.6cm, 0)$) edge node[left,at start,yshift=-0.2cm]{\small\xmark} ($(\curcord |- verifier.west) - (0cm, 0.1cm)$);

        \path ($(verifier.north) + (0.6cm, 0)$) -- ($(\curcord |- assumptioncheck.north east)$) -- coordinate[midway](correct) ($(assumptioncheck.north east) - (0.1cm, 0)$);
        \draw[red] ($(correct |- userinput) + (0, 0.3cm)$) node{Correct Code};
    \end{tikzpicture}}
    \caption{Overview diagram of our code generation system. The dashed lines between the user and Query Check and Assumptions Check indicate the user needing to review and approve or disapprove.}
    \label{fig:system-diagram}
    \end{minipage}
\end{figure}

Ansible programs, called \emph{playbooks}, are written in YAML~\cite{ansible.playbook}; an example is shown in Figure~\ref{fig:ansible-sample}.
A playbook is composed of a list of \emph{plays} which each specify configuration to perform on a particular set of remote systems; our example contains a single play.
Each play contains configuration information and a list of tasks to be performed in-order.
In our example, the configuration (lines~\ref{line:sample-play} through~\ref{line:sample-play-c}) includes the \texttt{name} field, which provides a natural language description of the play; the \texttt{hosts} to execute the play on (here ``servers'', the machine host names this refers to would be defined elsewhere); and the user to execute the play as on the remote machine (\texttt{become: true} causes execution as the user specified by \texttt{become\_user}).
Finally, \emph{tasks} are composed of configuration options and a \emph{module} which specifies the behavior of the task.
Task configuration can also include a \texttt{name} field, a \texttt{when} clause that causes the task to only execute if a condition is true, a \texttt{loop} clause that executes it on a list of elements, and so on.
Ansible modules are like library functions that define a particular set of operations that can be performed; they are distributed as \emph{collections} such as the built-in collection \texttt{ansible.builtin}. 
Our first task (lines~\ref{line:sample-task1} to~\ref{line:sample-task1-end}) invokes the builtin-in \texttt{package} module which is used to manage system packages; the arguments to this module include \texttt{name}, which specifies the packages, and \texttt{state}, which specifies whether to install or uninstall them.
In this task the \texttt{name} argument is a Jinja expression, that is evaluated at execution time to \texttt{apache2} if \mintinline{yaml}{ansible_os_family == 'Debian'} (the system is Debian based) and otherwise to \texttt{httpd}; this allows us to specify the correct package name based on the OS.
The other task (lines~\ref{line:sample-task2} to~\ref{line:sample-task2-end}) uses the \texttt{service} module to start the Apache server service.
Executing the playbook executes both tasks and so will install and then start Apache server and works on both Debian and RHEL.

\subsection{Errors in LLM-Generated Ansible}
While a detailed study of errors in LLM generated Ansible is beyond the scope of this work we briefly detail two subtle but important errors we observed regularly.
The first are incorrect package installations.
For example to install the Python package \textsf{numpy} on Red Hat Enterprise Linux (RHEL) we can install the \textsf{pip} package \texttt{numpy} or the \textsf{dnf} packages \textsf{numpy}, \textsf{python-numpy}, or \textsf{python3-numpy} but on Debian we must install the \textsf{apt} package \textsf{python3-numpy}.
We have seen LLM-generated code that uses \textsf{pip} on all systems, an incorrect package name on Debian, or \textsf{apt} on all systems; in all these cases the program fails to install \textsf{numpy} in some cases.
Errors like this are subtle as they require knowledge of Ansible, to understand the generated code, but also the underlying operating systems and package managers, to know the package names and package managers that can be used.
Our other example comes from permanently setting an environment; this is done by adding a line of the form \texttt{key=value} to the \texttt{/etc/environment} file.
Ansible has a module specifically for adding and replacing lines in files; to replace a line you specify a regular expression to search for and if a line matches it will be replaced.
In some generated code we observe a bug where it uses a regular expression like \texttt{\textasciicircum{}key} which will match and replace  any line with a variable which begins with \texttt{key}, such as \texttt{keys}, and therefore this error can result in unintentionally replacing other environment variables.
These observations offer guidance on the kinds of errors we need to detect to ensure correctness of LLM-generated Ansible.

\section{Problem Statement}\label{sec:problem-statement}
We are interested in providing correctness guarantees for code generated from a natural language description; to do this we need to first understand what it means for a program to be \emph{correct}.
Informally, what we mean is that the program acts as the user intends it to, but we cannot provide any formal guarantee of this if the user's intention is only in their mind and expressed, likely imperfectly, in natural language.
As a concrete example, consider a natural description like ``create a file at /some/path'', this is missing significant details like the contents, owner, group, permissions, and so on.
These missing details are important: the user likely has a vision that includes some of these details (for instance, they may want an empty file) and the system they want to configure may specify others (such as what users exist to own it).

To formalize this problem, let our target language be $\mathcal{T}$ (Ansible in our case).
The user has some \emph{intent} they want a program in $\mathcal{T}$ to satisfy; this is likely not a single program as there are generally many ways to achieve the same thing and there may be details the user does not care about, for instance if they want to clone a git repository, if git is already installed it is not necessary to install it but it does not hurt either.
Therefore, we can describe the user's intent as some set $\mathscr{U} \subset \mathcal{T}$; note importantly that we have \emph{no way} of directly testing membership in this set (the only option is to ask the user but this risks mistakes slipping through).
With this, though, we can now formalize our informal notion of correctness: a program $p \in \mathcal{T}$ is \emph{correct} if and only if $p \in \mathscr{U}$.

The user then writes some natural language description $\mathscr{d}$ to express their intent.
At this point, we still do not have any way to determine whether a program $p$ is correct as the only indication of the user's intent is $\mathscr{d}$ which lacks formal semantics.
To fix this, we need to introduce some formal specifications that allows us to test correctness.
For this, suppose that we have some formal specification language $\mathcal{S}$ (in our case the Ansible Formal Query Language discussed in Section~\ref{sec:fql-ansible}) and some semantics function $\den{\cdot} : \mathcal{S} \to \mathcal{P}(\mathcal{T})$, which defines the set of programs that satisfy a particular specification.
We also assume that given a program $p \in \mathcal{T}$ and specification $s \in \mathcal{S}$ we can decide whether $p \in \den{s}$.
Finally, suppose that we have some function $\mathscr{F}$ which converts a natural language description into a formal specification.

Because $\mathscr{F}$ takes a natural language description as input there must be instances where its output does not match the user's intent since the input may be ambiguous and $\mathscr{F}$ can only output one unambiguous formal specification.
This is inherent to any attempt to automatically formalize a natural language description.
The only solution to this problem is to show the generated specification $s = \mathscr{F}(\mathscr{d})$ to the user and ask them to check that it matches.
Once the user approves a specification we believe that the specification $s$ should satisfy $\den{s} \supseteq \mathscr{U}$ indicating that the specification contains the user's intent but may contain other programs as the user's intent may not be fully expressed in their description and thereby the specification.

Now, we can generate candidate programs $p$ until we find one that satisfies the specification, $p \in \den{s}$.
With this, we are close to a correct program but we need $p \in \mathscr{U}$, not just $p \in \den{s} \supseteq \mathscr{U}$.
With the information at hand: the natural language description $\mathscr{d}$ and the formal specification $s$, there is no way to check this.
We expect that $s$ is close to the user's intent and so if $p \not\in \mathscr{U}$ this should be because of how $p$ fills in under-specified details.
Therefore, we need to identify how $p$ differs from an \emph{unrefined} version of $s$, which does not make any particular choices for the under-specified details.
To do this, suppose we are able to compute an operation $p \setminus s$ which produces a high-level description of these choices.
Exactly what this operation and description look like will vary depending on the target language $\mathcal{T}$, but for Ansible these differences include details like the assumption that a program is installed or a particular group exists or that the program installs an unspecified program or sets a file owner when none was specified.
The result should be high-level so that it can be easily understood by the user and they can confirm whether or not the choices match their intent.
By ensuring that the program $p \in \den{s}$ we reduce the space of possible issues that the user must consider, but ultimately the task of determining whether $p \in \mathscr{U}$ is one that only the user can solve.

\section{Approach}\label{sec:approach}
We have proposed two key components that are the focus of this work: the \textbf{Formal Query Language}, our formal specification language $\mathcal{S}$, which expresses the user's intent in a formal manner, and the \textbf{Verifier} which, given a program $p$ and formal query $s$, determines whether $p$ satisfies $s$ ($p \in \den{s}$).
We envision that, in the future, these ideas could be incorporated into a full end-to-end system, like shown in Figure~\ref{fig:system-diagram}, that would guarantee the correctness of generated code as described in Section~\ref{sec:problem-statement}.

As described earlier, the Formal Query Language must be understandable to the user because they must check that it matches their intention.
We discuss the design considerations to satisfy this in Section~\ref{sec:fql-theory}.
For our ``Formalizer'' ($\mathscr{F}$) we propose to use an LLM; we note that the risk of mistakes by the LLM is not an additional problem here since there is no formalizer that can be correct all the time.
While we do not address the problem of generating formal queries in this paper, we believe it is suitable for LLMs, even though they are not trained on the formal query language, because in-context learning, where a few examples are provided to an LLM in the prompt, has shown efficacy at enabling LLMs to generate code in custom and domain-specific languages~\cite{lowresource.arxiv24}; LLMs have also been used before to formalize natural language descriptions into formal specifications such as postconditions~\cite{nl2postcond.fse24} and the generation of the formal query is not too dissimilar to the explanations of their work that LLMs often give.

Once the user is satisfied with a formal query, we can use it and their natural language description to generate programs (the ``Code Generator'' in Figure~\ref{fig:system-diagram}).
Note that the code generated by this step does not need to satisfy the formal query; in this work we actually do not use the formal query in this step and simply provide the natural language description to an LLM though we believe that providing the formal query could clarify the user's intent if there is ambiguity.

Finally, a generated program is checked against the formal specification by the ``Verifier''.
We discuss the role of the verifier in detail in Section~\ref{sec:verifier-theory}.
As Figure~\ref{fig:system-diagram} shows, the user may be asked to check the choices a program makes regarding the under-specification of the query ($p \setminus s$) via the ``Assumptions Check''.
In this work, we can identify such assumptions (i.e., our verifier calculus $p \setminus s$) but in our evaluation we accept all programs where $p \in \den{s}$.

\subsection{Design Considerations for a Formal Query Language}\label{sec:fql-theory}
As described earlier, to provide any formal guarantee of correctness we need a formal specification of what the program should do.
We do not need this formal specification to be complete, indeed program synthesis is often done on incomplete specifications, but do need this specification to be unambiguous.
As a concrete example, consider the following description for a desired Ansible program: ``copy a file from \texttt{/path/a} to \texttt{/path/b}.''
In the context of Ansible this could mean duplicating the file on the remote system, copying file \texttt{/path/a} from \emph{the Ansible controller} to \texttt{/path/b} \emph{on the remote system}, or from the remote system to the controller!
Any of these interpretations is conceivably correct, i.e., the query is ambiguous.
While we intended this example as the first interpretation, 
we find that all six LLMs we prompted \emph{only} generated code to copy a file from the controller to the remote system, potentially because this is the default mode implemented in the \texttt{copy} module.
This example raises a key design question:
\begin{itemize}
    \item[(Q1)] How do we ensure the formal specification matches the user's intent?
\end{itemize}

Section~\ref{sec:problem-statement} provides our answer: the user must check that the specification is correct.
We therefore need a language, that we call a \textbf{Formal Query Language}, that is designed to formalize the user's intent described in natural language.
Since the user must check the specification in this language, which we call a \textbf{Formal Query}, we want to make it easy and intuitive for the user and so the formal query should be close in syntax to natural language so that the user can read and understand it.
This means that our formal specification must be high-level and appear natural language-like.
The idea of natural language-like formal languages it not new, Controlled Natural Languages~\cite{cnl.cl14} have been explored before and it has been shown that such languages can be easy to understand~\cite{conceptualAuthoring.cl07} suggesting that can ensure that the specification is correct.
A further design question arises, though:
\begin{itemize}
    \item[(Q2)] How do we ensure the user does not incorrectly believe a query matches their intent?
\end{itemize}

For example, as described previously, on Debian the package name for numpy is \texttt{python3-numpy}, but a user may not know this off the top of their head and might believe a reasonable name like \texttt{numyp} is correct.
Our solution is that the formal query language must preserve certain high-level concepts, so rather than using package names we use \emph{package descriptions} such as ``numpy'' or ``apache server'' and use a trustworthy knowledge base to determine the correct package names and similar facts.
This use of high-level concepts and a knowledge base is part of why we are focusing currently on Domain-Specific Languages; the domain provides the high-level concepts that are needed and provides the information needed to construct the knowledge base.
We discuss our use of these design principles to build a formal query language in Section~\ref{sec:fql-ansible}.

\subsection{Program Verifier}\label{sec:verifier-theory}
Along with the Formal Query Language, the Program Verifier is the other key component to our approach.
The Verifiers's role is to take a generated program and a formal query and determine whether or not the program matches the \emph{user's intent}.
This is a significant challenge as there is no way to do it directly, we can only use the formal query and seek feedback from the user.
Let us consider the earlier example of cloning a git repository to exemplify this challenge.
In our experiments we find that some LLM generated programs for this will simply invoke ansible's \textsf{git} module (which requires git to be installed) and others will first install \textsf{git} before using it.
Both of these are reasonable: if the user's intent is to run the program on a machine with git installed they would likely be fine with either version (though there may be circumstances where they prefer or need the former) but if the user's target does not have git installed they need the second version.
A convenient solution for the Verifier would be to require that the formal query include a full description of such details, so specifying to install \textsf{git} if needed, but this would violate design question (Q2) as the user would need to think of these details and failing to do so, here simply failing to note whether they need to install git, could result in accepting incorrect code.

Thus, as described in Section~\ref{sec:problem-statement}, the Verifier's role is two-fold, to ensure that the program generally matches the query ($p \in \den{s}$) and therefore to detect certain errors like installing the wrong package or failing to create a file, \emph{and} the Verifier must also identify how the program fills in under-specified parts of the query ($p \setminus s$) and present them to the user who must check them.
The details of how a verifier does these things will depend highly on the target language and formal query language; we discuss our approach in Section~\ref{sec:state-calc-and-interp}.

\section{Formal Query Language and Verification of LLM-Generated Code for Ansible}\label{sec:ansible-impl}
We have implemented this design for Ansible in our system called \system{}.
In \system{} we have developed a formal query language and program verifier which together can be used (as shown in our Evaluation in Section~\ref{sec:evaluation}) to reliably identify correct and incorrect LLM generated Ansible code.
We describe the design of our Formal Query Language in Section~\ref{sec:fql-ansible}.
We then discuss the program verifier, which is built on a novel calculus we call the \textbf{State Calculus} that we use to define the semantics of Ansible (Section~\ref{sec:state-calculus}), a symbolic interpreter for this calculus (Section~\ref{sec:interpreter}), and a unification process that matches the expected and actual behaviors of a program (Section~\ref{sec:ansible-unifier}).
Finally, we explain how formal queries and Ansible programs are compiled into this calculus (Sections~\ref{sec:fql-compiler} and~\ref{sec:ansible-compiler}) and describe the formal query language's knowledge base (Section~\ref{sec:fql-kb}).

\subsection{Ansible Formal Query Language Design}\label{sec:fql-ansible}
As discussed already the role of the Formal Query Language is to provide a formal expression of the user's intent and must do so in a way so as to be understood, and reviewed, by the user.
We identified two key questions to consider in designing our Formal Query Language; our answers are:
\begin{itemize}
    \item[(A1)] The formal query language must be sufficiently high-level and interpretable that a user can easily understand whether a query matches their intent or not.
    \item[(A2)] Determining whether a query matches the user's intent must not rely on specialized knowledge (such as package names or assumptions made by Ansible modules).
\end{itemize}
These principles ensure that the user can verify a given query is correct and that errors do not go undetected by the user.

\begin{figure}
    \centering
    {\figsize
    \begin{align*}
        \text{query} ::=&\ \varepsilon
            \mid sentnc
            \mid sentnc\ .\ query \\[-.5mm]
        \text{sentnc} ::=&\ atom
            \mid atom\ ;\ sentnc
            \mid atom\ ;\ \textsf{and}\ atom
            \mid \textsf{if}\ cond\ \textsf{then}\ sentnc\ \textsf{otherwise}\ sentnc \\[-.5mm]
        \text{atom} ::=&\ verb\ desc\ args \\
        \text{cond} ::=&\ cond\ \textsf{or}\ cond
            \mid cond\ \textsf{and}\ cond
            \mid \textsf{os is Debian}
            \mid \textsf{os is RedHat}
            \mid desc\ args\ \textsf{exists}
            \mid \cdots \\[-.5mm]
        \vrb{verb} ::=&\ \textsf{create}
            \mid \textsf{delete}
            \mid \textsf{copy}
            \mid \textsf{move}
            \mid \textsf{enable}
            \mid \textsf{disable}
            \mid \textsf{install}
            \mid \cdots \\[-.5mm]
        \desc{desc} ::=&\ \varepsilon
            \mid \textsf{file}
            \mid \textsf{bashrc file}
            \mid \textsf{directory}
            \mid \textsf{user}
            \mid \textsf{group}
            \mid \textsf{numpy}
            \mid \textsf{apache server}
            \mid \cdots \\[-.5mm]
        \text{args} ::=&\ sep\ binds
            \mid sep\ val \\
        \sep{sep} ::=&\ \textsf{to}
            \mid \textsf{from}
            \mid \textsf{at}
            \mid \textsf{with}
            \mid \cdots \\[-.5mm]
        \args{binds} ::=&\ id = val[, binds] \\[-.5mm]
        \args{val} ::=&\ (id \mid string \mid int)^*
    \end{align*}}
    \vspace*{-2\baselineskip}
    \caption{Grammar of the Ansible Formal Query Language.}
    \label{fig:fql-grammar}
    \hspace{\parskip}
    {\figsize
    \begin{tabularx}{\textwidth}{|X|X|} \hline
        \textbf{Natural Language Description} &\textbf{Formal Language Query} \\ \hline
        Move the file at /scratch/file.txt to /home/user/data.txt
            &\vrb{move} \desc{file} \sep{from} \args{remote "/scratch/file.txt"} \sep{to} \args{remote "/home/user/data.txt"}
            \\ \hline
        Copy all .war files from /tmp on the remote machine into /backup on the controller
            &\vrb{copy} \desc{files} \sep{with} \args{glob="*.war"} \sep{from} \args{remote /tmp} \sep{to} \args{controller /backup}
            \\ \hline
        Create a Python 3.10 virtual environment at /user/home/venv and install numpy in it
            &\vrb{create} \desc{virtual environment} \sep{with} \args{python="3.10"} \sep{in} \args{/user/home/venv}; \vrb{install} \desc{numpy} \sep{in} \args{virtual environment} \sep{at} \args{/user/home/venv}
            \\ \hline
        Install zsh, set it as the user `dev''s default shell, and ensure its configuration file exists
            &\vrb{install} \desc{zsh}; \vrb{set} \desc{default shell} \sep{for} \args{user="dev"} \sep{to} \args{zsh}; if \desc{zsh configuration file} \sep{for} \args{user="dev"} not exists then \vrb{create} \desc{zsh configuration file} \sep{for} \args{user="dev"}
            \\ \hline
        Enable passwordless sudo for a `wheel' group and create a `ansible' user in that group
            &\vrb{enable} \desc{passwordless sudo} \sep{for} \args{group=wheel}; \vrb{create} \desc{user} \sep{with} \args{name=ansible} \sep{in} \args{supplemental groups=wheel}
            \\ \hline
    \end{tabularx}}
    \captionof{table}{Example Natural Language descriptions and Formal Language Queries for Ansible programming tasks. Colors and fonts of the formal queries correspond to the non-terminals in Figure~\ref{fig:fql-grammar}.}
    \label{tab:fql-examples}
\end{figure}

From these principles we have developed a Formal Query Language for Ansible; a fragment of the grammar is shown in Figure~\ref{fig:fql-grammar} and a number of examples are shown in Table~\ref{tab:fql-examples}.
Towards principle (A1), queries are designed to mimic the structure of natural language; they are sequences of sentences which are, in turn, sequences of basic atomic queries or a conditional sentence like the fourth example, where \texttt{if zsh configuration file for user="dev" not exists} allows the query to specify what should happen in different circumstances.
Atomic queries specify simple imperative actions and are defined by a verb (such as \texttt{move} in the first example or \texttt{copy} in the second), a description (like the package description \texttt{zsh} or file desciption like \texttt{zsh configuration file} in the fourth query), and finally additional arguments defined either by a separator and a value (such as \texttt{at /user/home/venv} in the third query) or a separator followed by key-value argument pairs (like \texttt{with glob="*.war"} in the second example).
Separators act both to separate the description from the arguments and when followed by a value act as the variable name, so \texttt{to remote "/home/user/data.txt"} in the first example sets the \texttt{to} argument (specifying the destination) to the given path.

In following the (A2) principle, our design reduces the effort required by users to verify a query by preserving high-level descriptions, such as \texttt{numpy} in the third example (rather than using the actual package name), \texttt{zsh configuration file for user="dev"} in the fourth (rather than using the actual path), and \texttt{passwordless sudo} in the fifth (rather than describing the changes needed to the sudoers file).
Details on how we achieve this are discussed in Section~\ref{sec:fql-kb}.

Returning to the example Ansible program in Figure~\ref{fig:ansible-sample}, which installs and starts Apache server on Debian and RHEL, a formal query for this problem is: \texttt{if os is Debian or os is RedHat then install apache server; and start apache server service}.
This example illustrates that our formal queries are significantly easier to understand than the Ansible code, are close to natural language in structure while still being precise, and are easier to verify the correctness of.

\subsection{State Calculus and Interpreter}\label{sec:state-calc-and-interp}
In order to verify the correctness of Ansible programs we compare the behavior of the program and expected behavior defined by the formal query.
To do this, we have developed the \textbf{State Calculus} which provides a clean mechanism to describe the semantics of Ansible programs (Section~\ref{sec:state-calculus}).
Given an Ansible program and a formal query, both compiled into the State Calculus (see Sections~\ref{sec:fql-compiler} and~\ref{sec:ansible-compiler} for details), we make use of a symbolic interpreter to identify their behaviors (Section~\ref{sec:interpreter}) which we compare to determine whether the program satisfies the query (Section~\ref{sec:ansible-unifier}).

\subsubsection{State Calculus}\label{sec:state-calculus}
Programs in languages like Ansible and Bash are executed mostly for side-effects in the form of changes to the system state.
We can model such programs as executing with some current system state they read information from (such as whether a file exists) and make modification to.
This gives rise to our \textbf{State Calculus}, which is a simple calculus designed to model programs that operate in this manner.

\paragraph{States}
We define states generically, so as to be applicable to Ansible, other languages, and even other domains.
States, denoted by $\sigma$, are defined in Figure~\ref{fig:states-def}; formally, they are a pair of functions, but we can think of it as a set of two kinds of values: attributes and elements.
Attributes, described by \emph{attribute labels} from some set $\mathcal{A}$, map to a value $\nu$; these attribute values are used to describe information in the state that has a value.
For example, we might have an attribute \textsf{os} whose value describes the operating system we are running on or an attribute \textsf{file\_mode} describing the file permissions of a certain file.
Elements are used to model the existence or non-existence of certain objects in the state, such as files or users.
An element is specified as an \emph{element label} from $\mathcal{E}$ and a value $\nu$; for instance, if $\mathsf{file} \in \mathcal{E}$ then $\mathsf{file}(\texttt{/etc/passwd})$ is an element referring to the \texttt{/etc/passwd} file.
Where attribute labels identify a piece of information, an element label identifies a class of objects and when applied to a value it identifies a particular object.
In our states, elements map to either the special value $\bullet$, indicating the element does not exist, or to a \emph{nested state} $\sigma$, which allows us to apply attributes on an element.
For example, the $\mathsf{file}(\texttt{/etc/passwd})$ element can map to a state with attributes describing the owner, the file permissions, and so on.
As an example, a state with $\mathsf{file}(\texttt{/etc/passwd}) \mapsto [ \mathsf{owner} \mapsto \texttt{root}, \mathsf{mode} \mapsto \texttt{0644}], \mathsf{file}(\texttt{/root/notes.txt}) \mapsto \bullet$ indicates that the \texttt{/etc/passwd} file exists, is owned by root, and has permissions 0644, and the file \texttt{/root/notes.txt} does not exist.

\begin{figure}
    \centering
    \begin{subfigure}[b]{0.32\textwidth}
        \centering
        {\figsize
        \begin{align*}
            \sigma &\equiv (\mathcal{A} \to \nu) \times (\mathcal{E} \times \nu \to \tau) \\
            \tau &::= \sigma \mid \bullet \\
            \mathcal{E} &: \text{the set of element labels} \\
            \mathcal{A} &: \text{the set of attribute labels} \\
            \nu &: \text{the set of values}
        \end{align*}}
        \caption{Formal Definition of States}
        \label{fig:states-def}
        {\figsize
        \begin{align*}
            \mathscr{s} &\equiv (\mathcal{A} \rightharpoonup \nu) \times (\mathcal{E} \times \nu \rightharpoonup \tau)
        \end{align*}}
        \caption{Symbolic State Definition}
        \label{fig:abs-state-def}
    \end{subfigure}
    \hfill
    \begin{subfigure}[b]{0.67\textwidth}
        \centering
        {\figsize
        \begin{align*}
            p ::=&\ s^* \\
            s ::=&\ x \gets e
                \mid x \gets \alpha(e)
                \mid x \gets \textbf{get}\ a
                \mid \textbf{add}\ c
                \mid \textbf{ret}\ e
                \mid \textbf{fail} \\
                \mid&\ \textbf{if}\ e\ \textbf{then}\ p\ \textbf{else}\ p
                \mid \textbf{exists}\ l\ \textbf{then}\ p\ \textbf{else}\ p \\
                \mid&\ \textbf{case}\ e\ \textbf{of}\ \textbf{L}(x) \Rightarrow p\ \textbf{R}(x) \Rightarrow p \\
                \mid&\ \textbf{foreach}\ x\ \textbf{in}\ e\ \textbf{do}\ p \\
            e ::=&\ x
                \mid \ell
                \mid \langle e , e \rangle
                \mid \rho(e)
            \quad\quad a ::= \mathcal{A}
                \mid t . \mathcal{A}
            \quad\quad c ::= a \Leftarrow e
                \mid l \\
            l ::=&\ t
                \mid \overline{\mathcal{E}(e)}
                \mid t . \overline{\mathcal{E}(e)}
            \quad\quad t ::= \mathcal{E}(e)
                \mid t . \mathcal{E}(e) \\
            x :&\ \text{variables}
            \quad\quad \alpha: \text{stateful functions}
            \quad\quad \ell: \text{literals} \\
            \rho :&\ \text{pure functions}
        \end{align*}}
        \caption{Syntax of the State Calculus}
        \label{fig:state-calculus-syntax}
    \end{subfigure}
    \caption{Description of the State Calculus}
    \label{fig:state-calculus}
\end{figure}

\paragraph{Syntax}
The grammar of the State Calculus is shown in Figure~\ref{fig:state-calculus-syntax}, much of which is fairly standard.
A program $p$ is a sequence of statements $s$ which are a mix of standard and unique forms.
These begin with variable assignment $x \gets \cdots$ where the right hand side is a pure expression $e$ (here pure means they do not read or modify the state), or special forms $\alpha(e)$ or $\textbf{get}\ a$.
 $\alpha(e)$ represents the invocation of a \emph{stateful function} $\alpha$. 
For Ansible, these stateful functions include the Ansible modules but can be any function that can be defined by a program in the State Calculus.
The other unique form, $\textbf{get}\ a$, is used to read the value of the attribute $a$; which is either an attribute label from $\mathcal{A}$ or an attribute label on an element $t . \mathcal{A}$, where the element $t$ can be a single element, $\mathcal{E}(e)$, or a sequence of nested elements.
For example, we load the owner of a file with $x \gets \textbf{get}\ \textsf{file}(\texttt{/path}).\textsf{owner}$.
The next statement form, $\textbf{add}\ c$, is used to modify the state by assigning a value to an attribute, $a \Leftarrow e$, or adding an element $l$.
As an example, the statement $\textbf{add}\ \textsf{file}(\texttt{/path}).\textsf{owner} \Leftarrow \text{``user''}$ sets the owner of the \texttt{/path} file.
We can also specify a \emph{negated element}, $\overline{\mathcal{E}(e)}$, to remove an element (i.e., map it to $\bullet$).
So $\textbf{add}\ \overline{\textsf{file}(\texttt{/path})}$ represents deleting the file.
The next two forms, return and failure, are standard.

The remaining four statement forms are control flow. if-then-else statements, for-each loops, and case statements are all standard and behave as expected.
The fourth is the \emph{exists-then-else} statement which acts like an if-then-else statement but the condition tests whether an element $l$ exists (does not map to $\bullet$).
The exists-then-else statement allows us to read elements from the state and serves as a complement to \textbf{get} which reads attributes.
Finally, the expression forms in the calculus are mostly standard, with variables $x$, literal values $\ell$, pairs $\langle e, e \rangle$, and finally function calls to \emph{pure functions} $\rho(e)$, which include built-in operators.

\subsubsection{Symbolic Interpreter}\label{sec:interpreter}
To determine the behavior of programs in the State Calculus, we develop a symbolic interpreter that determines the behavior on all possible initial states.
The result of the interpreter is a set of \emph{symbolic initial-final symbolic state pairs} that reflect the behavior of the program under different initial states.
To do this, the interpreter tracks both an initial state and the cumulative modifications made to the state (which we often imprecisely refer to as the ``current state'').
These initial and cumulative states are defined as symbolic states $\mathscr{s}$, as defined in Figure~\ref{fig:abs-state-def} where the attribute and element maps are partial functions to record only the attributes and elements that are relevant to the program.
The design and implementation of the interpreter are mostly straightforward.
The main complexity is handling symbolic values and the state which we discuss below.

The three constructs dealing with the state are \textbf{get}, \textbf{add}, and \textbf{exists}.
Of these, \textbf{add} is the simplest, we simply record the addition into the current state.
For \textbf{get} and \textbf{exists} there are two scenarios, either we can determine the result based on the existing state or the result depends on undefined parts of the state.
The first scenario is the simplest; for a \textbf{get} we search the state for the attribute and find a value for it in the current or initial states and we simply use that value (if we find the attribute in both we use the value in the current state as this is the up-to-date value).
Similarly, for \textbf{exists}, we find the element bound in the current or initial state (again preferring what we find in the current state) and take the appropriate branch.

The other scenario is that the attribute or element being read is undefined.
In this case, to interpret the statement, we need to define this missing piece.
For \textbf{get} all we need to do is add the attribute to the state and give it a symbolic value representing that it can have any value.
The one complexity is determining whether to add it to the initial or current state; in general we add it to the initial state (since generally the properties attribute represent depend on the initial system configuration) but if the attribute is nested on an element not present in the initial state we add it to the current state instead.
As examples, the first case occurs when a file existed originally and we read the owner which is some (unknown) value while the latter case occurs when a file was created and we read the owner which is some unknown value as it was not explicitly set.
Handling \textbf{exists} is similar, except we have to handle both cases: where the element exists or not (is bound to $\bullet$).
To do this, we will separately update the state for both cases and interpret the \textbf{then} and \textbf{else} branches respectively.
For updating the state we always update the initial state since elements cannot be unspecified like attributes can and so much always originate from the initial state.

The last interesting pieces of the interpreter are how symbolic values are handled by if-then-else, case, and foreach statements.
For if-then-else and case statements when the condition or scrutinee is symbolic we must interpret both branches, like we did for \textbf{exists}.
When we do this we record the value assumed for the symbolic value (i.e., true or false for a condition based on the branch we take) so the value can be reused again later if needed.
Handling for-each loops over symbolic values is more challenging.
For a concrete list, we simply interpret the body of the loop once for each element of the list, but for a symbolic value we do not know the elements but still want to execute the loop body so that the side-effects of the loop are performed.
To do this, we create a new symbolic value to represent the elements of the list and interpret the loop body \emph{once} using this symbolic value as the element.
We then record that this symbolic value represents all elements of the list, information which is used in our verifier, as described below.
While this approach to handling symbolic lists is imprecise, loops in Ansible are simple enough that this approach captures their general behavior.

\subsubsection{Program Verifier}\label{sec:ansible-unifier}
Finally, to verify that the behaviors of the program and query match we compare the results of interpreting them; we call this component the \emph{unifier}.

\paragraph{Basic Operation}
By compiling the formal query and Ansible program to the State Calculus (Sections~\ref{sec:fql-compiler} and~\ref{sec:ansible-compiler}) we can run the symbolic interpreter to produce a set of symbolic initial-final state pairs for each (we refer to these as the query's and program's state pairs, respectively).
To verify that the program's behavior matches the formal query's we need to ensure that they behave the same under the same initial conditions; the initial conditions are represented by the initial state and the behavior of the program (i.e., what it side-effects is has) by the final state.
To do this, for each of the formal query's state pairs $s$ we need to find one of the program's state pairs $q$ such that $q$'s initial state is \emph{consistent} with $s$'s and their final states are the same; such a pair indicates that under the initial conditions described in $s$, the program behaves the same as the query.
Note that we only require $q$'s initial state to be \emph{consistent} with $s$'s, rather than the same; this is because the program is not required to rely on all the assumptions that the query does.
For example, a query like \texttt{if os is Debian install git} will have an initial state that assumes the OS is Debian, but the program may not ever check the OS since it is not needed (since the package name is always the same) and so that assumption may not appear in the initial states of the program.
The final states we do require to be the same, since the program must behave the same as the query, and if anything is missing from the final state it means the program did not behave as expected.

Because the interpreter's results often involve symbolic values, this process also involves some unification-like techniques where symbolic values may be unified or concretized.
As an example, based on a problem in our benchmark suite, suppose we have a query like \texttt{if os is Debian then create file "Debian". if os is RedHat then create file "RedHat".}, when we interpret this query these two file names appear directly in the resulting state pairs but a generated program might just use Ansible's OS family variable as the filename because it has the appropriate value, and so in the program's resulting state pairs the specific names ``Debian'' and ``RedHat'' are not used, instead a symbolic value is used.
This program works, and so to handle this we need to be able to concretize symbolic values in the program's state pairs into concrete values to match the query's.
We can also need to unify symbolic values between the query's and the program's state pairs, for instance when both have a file with some unknown contents; note that we will only unify a symbolic value which represent elements of a symbolic list with another value that represents the value of another symbolic list (and that we can unify their symbolic lists), this ensures that we preserve the knowledge that they represent many values at once.
Note that we cannot concretize symbolic values in the query's state pairs to match a concrete value in a program's state pair, because a symbolic value in the query's state pair indicates that the behavior is expected for all possible values and so the program should work for all values as well.

\paragraph{Handling under-specified queries}
As discussed in Section~\ref{sec:verifier-theory} it is often the case that a formal query is an under-specification and this means that a generated program may be valid even if makes additional assumptions (like that a certain program is already installed) or performs additional actions (like install an additional program) compared to the formal query.
In addition to allowing this, the unifier also needs to accumulate these assumptions and actions so they can be shown to the user (this is $p \setminus s$ as described in Section~\ref{sec:problem-statement}).
To do this, we have to change how we check initial states for consistency and loosen the requirement on final states.
For initial states, we already allow programs to have fewer assumptions than the query and now we need to also allow the program to have additional assumptions; thus consistency is now just that if the program's and query's initial states contain a common assumption then it is consistent (i.e., they do not disagree about the existence of an element or the value of an attribute); any assumptions in the program's initial state not present in the query's initial state are reported as additional assumption.
Similarly, for final states, we had previously assumed that the program's and query's final states must be the same but now we permit the program's to include additional actions (i.e., additional elements or attributes) and we report any of these as additional actions; importantly it is still the case that all actions in the query's final state must be matched exactly in the program's final state to ensure the program performs the actions expected of it.

\subsection{Compiling Formal Queries to the State Calculus}\label{sec:fql-compiler}
Compiling a formal query to the State Calculus is mostly straightforward; a query is a sequence of atomic queries, the generated code of which are composed together in sequence.
Conditional queries are compiled into if-then-else statements after compiling the condition; boolean combinations of conditions are compiled to pure functions that perform these operations, and the base conditions, such as checking the OS or whether a file exists, are compiled into specific code for that purpose such as reading an attribute that represents the OS and checking its value.

Compilation of atomic queries is based on the verb and description.
For example, creating a file and a directory generate different code, but creating a file always generates the same code regardless of the description of the file: we add a file element with the appropriate path and set attributes based on the arguments provided.
Similarly, installing a package is always the same: we add an element indicating the package's presence and set attributes like the package manager.
Other atomic queries are similarly straightforward.
Currently the set of verbs we support, and the outline of how to compile them, are built-in to the compiler but we believe these could be made extensible easily, such as through our knowledge base which is already used for handling descriptions (see Section~\ref{sec:fql-kb}).

\subsection{Compiling Ansible Programs to the State Calculus}\label{sec:ansible-compiler}
To compile Ansible programs into the State Calculus we use a relatively simple compiler to translate playbooks.
This compiler relies on definitions of Ansible modules (modules are like library functions that Ansible programs are built from) that provide type information and the semantics as a program in the State Calculus.
To facilitate these definitions, we develop a \emph{module definition language}.
We describe the Ansible compiler and module description language in turn.

\subsubsection{Ansible Playbook to State Calculus Compiler}
The actual compilation of an Ansible program into the State Calculus is surprisingly simple.
We generate code for each task in each play in sequence and so the core of the procedure is just generating code for an individual task.
The module itself becomes an invocation of a stateful function for that module and other features, like conditional execution, storing the returned value, and iterating over a list are straight-forward to handle.
The main complexity then is constructing the input as this relies heavily on type information to ensure the arguments used are appropriate and have the correct types.
This compilation must also identify built-in Ansible variables, such as those providing information on the OS, and convert them into reading appropriate attributes.

\subsubsection{Module Description Language}
As mentioned above, the definitions of modules and their types are essential to compiling and interpreting Ansible programs.
It would be ideal if this information could be extract from an existing source, such as a module's documentation or (Python) implementation but unfortunately neither is suitable.
While some type information could be automatically extract from the documentation important details like when two arguments are mutually exclusive are only described in natural language.
As for the implementations, the code is too low-level and complex to extract the core behavior we are looking for.
We therefore designed the module description language to facilitate defining the behavior and types of modules.
The language includes features for specifying Ansible's module argument, which include optional arguments, some with default values, and sets of arguments where exactly one is required.
Many Ansible modules also have arguments that must take a value from some list of values which we represent with custom enum types.
The module description language also enables defining new elements and attributes as needed for the modules being specified.

Figure~\ref{fig:copy-module-def} shows a simplified definition for the Ansible copy module, the omitted details deal with the many additional arguments (like file owner and mode) the module supports and details for copying directories.
This demonstrates the features we have described.
Arguments are defined on lines~\ref{line:required-var} (\texttt{dest} is a required argument), \ref{line:required-var-options} (exactly one of \texttt{src} and \texttt{content} is required), and~\ref{line:opt-var} (\texttt{remote\_src} is optional).
When arguments are optional, or one of several options, we can check if a particular one was provided like on line~\ref{line:var-provided-check} where
in the then-branch \texttt{src} was provided and can be used (see line~\ref{line:state-updates}) while in the else-branch \texttt{content} was provided and can be used (see line~\ref{line:else-branch-update}).
Similarly, we check whether \texttt{remote\_src} was provided on line~\ref{line:provided-remote-src}.
The module's behavior relies a lot on the state, for instance line~\ref{line:if-exists} ensures the source file exists, and lines~\ref{line:state-updates} and~\ref{line:else-branch-update} set the contents of the destination file based on the source file and \texttt{contents} argument, respectively.

\begin{figure}
    \centering
\begin{lstlisting}[style=module]
module ansible.builtin.copy -> bool {
    (dest: path)!$\label{line:required-var}$!
    (src: path | content: string)!$\label{line:required-var-options}$!
    if provided src {!$\label{line:var-provided-check}$!
        [remote_src: bool]!$\label{line:opt-var}$!
        let src_loc = provided remote_src!$\label{line:provided-remote-src}$!
            ? (remote_src ? file_system::remote : file_system::controller)
            : file_system::controller;
        if exists file(src, src_loc) {!$\label{line:if-exists}$!
            file(dest, file_system::remote).content = file(src, src_loc).content;!$\label{line:state-updates}$!
        } else {
            return false;
        }
    } else {
        file(dest, file_system::remote).content = content;!$\label{line:else-branch-update}$!
    }
    return true;
}
\end{lstlisting}
    \caption{A simplified module description for \texttt{ansible.builtin.copy}}
    \label{fig:copy-module-def}
\end{figure}

We have designed the module language to make it easy to write specifications for modules, since a specification is needed for every supported module.
Towards this goal, it gives significant flexibility in deciding the level of detail tracked in the state.
For example, we model file systems at a rather coarse level, like shown in Figure~\ref{fig:copy-module-def}, by just tracking the path and system we are referring to, but this approach does not model hard-links, where two files share the same inode, correctly; if we care about such details
the module language provides the flexibility to model them, perhaps by defining an inode attribute for files and having an \textsf{inode} element that maps an inode number to its contents.
We have also found that the language makes it easy to add features of a module as needed by adding the extra arguments and defining their uses as needed.

\subsection{Formal Query Language Knowledge Base}\label{sec:fql-kb}
As described briefly earlier, we make use of a Knowledge Base in compiling formal queries into the State Calculus.
This Knowledge Base is essential in compiling high-level descriptions of packages, files, and so on from the Formal Query Language into the State Calculus.
This is essential to the design of the Formal Query Language as it ensures that users do not accept incorrect but plausible sounding queries, which as we have described previously is an issue with LLM-based code generation.
The knowledge base provides a set of functions that lower these high-level descriptions into lower-level ones appropriate for the State Calculus.
For example, a \texttt{pkgDef} function takes in a package description and context information (including what OS is being targeted) and returns the appropriate package name and package manager; for instance, if the package description is ``apache'' or ``apache server'' and the OS being targeted is Debian-based it will return the package name \texttt{apache2} and the \texttt{apt} package manager while if the OS is RHEL it will return the name \texttt{httpd} and \texttt{dnf} package manager.
Similarly, we have a \texttt{fileDef} function which given the description ``zsh configuration file for user=foo'' or ``zshrc file for user=foo'' will respond that this refers to the \texttt{.zshrc} file in foo's home directory on the remote machine.

In our implementation, we define an interface for the knowledge base which contains these various functions we need and then parameterize our compiler by an instance of this interface.
This makes it easy to replace and modify the knowledge base in the future.
Currently, our knowledge base has been hand curated but it can be easily extended in the future and we are interested in exploring how to automate the construction by leveraging existing trustworthy knowledge sources, such as publicly available lists of packages and documentation such as Linux manual pages.

\section{Evaluation}\label{sec:evaluation}
To evaluate our approach, we study \system{}'s ability to differentiate correct and incorrect LLM-generated code.
We present a benchmark suite of natural language descriptions and formal queries for common Ansible tasks (Section~\ref{sec:data-set}), 
describe our code-generation methodology and the quality of the LLM generated code (Section~\ref{sec:llm-code-gen}), and finally discuss the system's accuracy (Section~\ref{sec:verifier-accuracy}).

\subsection{Benchmark Suite}\label{sec:data-set}
For our evaluation, we need natural language descriptions and formal language queries for tasks to be done in Ansible.
Prior work has used the \texttt{name} fields from tasks and plays as natural language descriptions~\cite{ansiblecodegen.arxiv23} but in our experience these fields are generally quite short and often lack important details, like operating systems and paths.

Given the lack of a suitable existing benchmark suite, we develop our own by identifying common tasks based on top posts on StackOverflow and the Ansible Forum.
In addition, we add tasks based on examples found in Ansible Galaxy, a large repository of Ansible code, some interesting variations on other tasks, and tasks that highlight challenges described earlier.
For each benchmark we develop a natural language description, a formal language query, and a reference solution.
We have produced a set of 21 benchmarks which require an average of $1.67$ tasks in our reference solutions and a maximum of $4$ tasks, putting them in line with the size of playbooks generated by prior work~\cite{ansiblecodegen.arxiv23}.
The natural language descriptions and formal queries included in our benchmark suite are provided in Appendix~\ref{app:benchmark-suite}; the examples in Table~\ref{tab:fql-examples} are taken from the suite.
The benchmark suite includes a mix of tasks involving file, user, and package operations.

Correctness checking requires a validated ``ground-truth'' program version we can use for comparisons.
To do this we use tests because, even though they are imperfect, no other solution was feasible: as described in Section~\ref{sec:llm-code-gen} we have 1260 programs, making manual inspection infeasible. We have done our best to craft strong tests that address subtle aspects of our problems.
To test generated programs, we run them on a set of virtual machines along with setup and testing scripts which determine whether the program behaved as expected.
We use multiple tests to test the behavior under different initial conditions, such as for problem 6 where a file should only be created if it does not already exist.
We use VMs running Debian 12.11.0, Ubuntu 24.04.2, and Red Hat Enterprise Linux (RHEL) 9.6; for any problem that does not specify an operating system we test the program on each system and for problems that specify certain operating systems we run it on those systems.
Between each program execution we reset the VMs from a snapshot to ensure that a program cannot impact the behavior of another.
A program is judged to be correct only if it passes all tests on all systems it was tested on.

\subsection{LLM Code Generation}\label{sec:llm-code-gen}
To generate a diverse set of programs to test \system{}, we have selected six LLMs and since the LLMs are nondeterministic we prompt each for ten programs for each problem to further increase the diversity of our program set, resulting in a total of 1,260 programs.
The prompt we use is shown in Appendix~\ref{app:prompt}; it gives basic instructions and an instruction not to make use of shell commands, such as the \texttt{shell} and \texttt{command} modules which allow the execution of arbitrary shell code, because we do not currently support these modules and their use is unnecessary in our benchmarks and therefore discouraged~\cite{ansible-lint.shell}.
Despite this instruction we find 130 instances of generated code containing them; these programs are identified as incorrect by our verifier even though in some cases they do pass our tests.

In our experiments we use five open source-models which perform well on other code-generation benchmarks: Deepseek Coder 6.7b~\cite{deepseek-coder.arxiv24}, Granite 8b Code~\cite{granite.arxiv24}, Qwen2.5 Coder 3b~\cite{qwen25-coder.arxiv24}, Llama 3.1 8b~\cite{llama-3.arxiv24}, and Starcoder 2 15b~\cite{starcoderv2.arxiv24}.
We also use a state-of-the-art closed-source model GPT-4o~\cite{gpt-4.arxiv24}.
We have observed that the models do not always generate full playbooks, in some cases models only generate a list of tasks; to address this, and ensure the playbooks work with our testing setup, we perform post-processing to identify and extracts key elements and insert them into a template.
In this process, we fail to process 42 (3.3\%) of the generated programs due to syntactic issues.
We also identified 5 programs (0.4\%), all solutions to problem 21, which do not terminate because they invoke a shell command which requires user input.
Of the remaining programs, we find that 596 (47.3\%) fail to execute and 226 (18.0\%) execute but fail the tests.
In addition, the solutions to problem 18 cannot be executed (they attempt to access a URL that does not exist) but through manual inspection we identified only 1 of the 58 syntactically valid responses is correct.
This means that in total, of the 1260 generated programs only 334 (26.5\%) are correct.

We break down the performance of each model in Table~\ref{tab:model-errors}.
GPT-4o performs significantly better than the other models, generating 108 correct programs, or 51.4\%, in contrast to only 21.5\% from the open source models.
This difference is most likely because the open source models are much smaller than GPT-4o.
Within the open source models, Deepseek, Granite, and Starcoder all achieve slightly over 25\% while Llama and Qwen only achieve around 15\%.
Qwen is the smallest model with only 3 billion parameters and Llama is the only non-code model included in our evaluation, which explains their low performance and may particularly explain Llama's significantly higher than average rate of syntax errors.

\subsection{Verifier Accuracy}\label{sec:verifier-accuracy}
To determine how well \system{} works at differentiating correct from incorrect code, we run it on each on the programs generated by the LLMs using our hand-written formal queries.
A perfect result would be that it identifies exactly the 334 correct programs as correct and all other programs as incorrect but some programs that are correct are rejected and some programs that are not correct are accepted.
Our overall results, by benchmark program, are shown in Table~\ref{tab:eval-results}.

\begin{table}
    \centering
    \begin{minipage}{.38\textwidth}
    \centering{\figsize
    \begin{tabular}{|p{0.7in}|c|c|c|c|} \hline
        \multirow{2}{*}{\textbf{Model}} &\multicolumn{4}{c|}{\textbf{Error}} \\
        &\textbf{S} &\textbf{E} &\textbf{T} & \textbf{C} \\ \hline
        Deepseek Coder 6.7b &  0 & 87 & 70 & 53 \\ \hline
        GPT-4o &  2 & 65 & 35 &108 \\ \hline
        Granite-8b Code &  0 & 89 & 66 & 55 \\ \hline
        Llama 3.1 8b & 22 &115 & 40 & 33 \\ \hline
        Qwen2.5 Coder &  4 &155 & 20 & 31 \\ \hline
        Starcoder 2 15b & 14 & 90 & 52 & 54 \\ \hline
        \textbf{Total} &42 &601 &283 &334 \\ \hline
    \end{tabular}}
    \caption{Summary of the error types of code generated by model. Each model produced a total of 210 programs.
    The columns count the number of programs with \textbf{\underline{S}}yntax Errors, \textbf{\underline{E}}xecution Errors, \textbf{\underline{T}}est Failures, and programs that are \textbf{\underline{C}}orrect.
    The programs which do not terminate are counted as Execution Errors and programs manually identified to be incorrect code are counted as Test Failure errors.}
    \label{tab:model-errors}
    \end{minipage}
    \hfill
    \begin{minipage}{.6\textwidth}
    \centering{\figsize
    \begin{tabular}{|c|c|c|c|c|} \hline
        \multirow{2}{*}{\#} &\multicolumn{2}{c|}{\textbf{Correct Programs}} &\multicolumn{2}{c|}{\textbf{Incorrect Programs}} \\
         &\textbf{Accepted} &\textbf{Rejected} &\textbf{Accepted} &\textbf{Rejected} \\ \hline
         1 &55 & 0 & 5 & 0 \\ \hline
         2 &56 & 0 & 0 & 4 \\ \hline
         3 & 5 & 2 & 0 &53 \\ \hline
         4 & 0 & 0 &49 &11 \\ \hline
         5 & 5 & 4 & 0 &51 \\ \hline
         6 &11 & 0 & 4 &45 \\ \hline
         7 & 3 & 0 & 0 &57 \\ \hline
         8 & 2 & 0 & 0 &58 \\ \hline
         9 &37 & 7 & 0 &16 \\ \hline
        10 & 5 &22 & 0 &33 \\ \hline
        11 &42 & 0 & 2 &16 \\ \hline
        12 &11 & 0 & 3 &46 \\ \hline
        13 & 0 & 0 & 0 &60 \\ \hline
        14 & 0 & 6 & 0 &54 \\ \hline
        15 & 0 & 0 & 0 &60 \\ \hline
        16 &14 & 8 & 0 &38 \\ \hline
        17 &24 & 0 & 2 &34 \\ \hline
        18 & 0 & 1 & 2 &57 \\ \hline
        19 & 7 & 0 & 3 &50 \\ \hline
        20 & 0 & 7 & 0 &53 \\ \hline
        21 & 0 & 0 & 0 &60 \\ \hline
        \textbf{Total} &277 &57 &70 &856 \\ \hline
    \end{tabular}}
    \caption{Evaluation Results for the Benchmark Suite. For each benchmark we have 60 samples for a total of 1260 programs.}
    \label{tab:eval-results}
    \end{minipage}
\end{table}

Of the 334 correct programs \system{} accepts 277 of them (82.9\%) and of the 926 incorrect programs \system{} rejects 856 of them (92.4\%).
In total, \system{} took about 70 seconds to process the 1260 programs.
In this process it detects many kinds of errors, including all of the errors we have used as examples throughout this paper, as well as many errors involving using incorrect modules or module names, providing invalid arguments or missing necessary ones, providing invalid argument values, and so on.
For the remainder of this section, we discuss the cases where \system{} is incorrect and how these issues can be addressed.

Of the 57 correct programs that \system{} rejects, 18 are because they rely on features we do not support yet; 9 of these rely on features like \texttt{block} constructs and handler, which are easy to support, while other 9 rely on shell commands which will be more difficult to support, as it requires a specification of the behavior of shell programs, we believe this is possible and discuss the possibility in Section~\ref{sec:generality}.
The other 39 programs come from problems: 9, 10, 16, and 20; these all share a common issue: the knowledge base is only allowed to produce one response and these programs used a correct value (or approach) that did not match the one the knowledge base provided.
For example, in problem 9 we cloning a git repository and the programs use the address \texttt{https://github.com/ansible/ansible} while knowledge base produced this same address with \texttt{.git} at the end.
Similarly, problem 10 which sets an environment variable, expects that we insert the line be \texttt{LC\_ALL=C} while the line \texttt{LC\_ALL="C"} works as well.
Problem 16's issue is the same, the knowledge base provides the path \texttt{/bin/zsh} for zsh but the path \texttt{/usr/bin/zsh} work as well.
Again, for problem 20 where we enable sudo for a group, the knowledge base provides a particular line to be added to the \texttt{/etc/sudoers} file but multiple acceptable lines exist and it can be added to other files as well.
The simple solution to these is to allow the knowledge base to return multiple values, which would fix these issues, but unfortunately the implementation is not quite as trivial, while we believe it can be resolved it is likely to be rather technical and may require changes to our handling of symbolic values.
Thus, all of these incorrect rejections are likely fixable, though require varying levels of effort to fix.

Of the 70 incorrect programs that \system{} accepts we find that 68 are the result of the queries being under-specified and that we fact that we are accepting all programs that satisfy the query (i.e., without a human verifying $p \setminus s$).
of the other two programs, one is the result of an unhandled feature, Shell plugins, which in compiling the Ansible to the State Calculus we simply ignore but causes Ansible to fail to execute it; handling Shell plugins in our compilation of Ansible would immediately address this issue.
The other program is due to an undocumented feature in the lineinfile module~\cite{ansible.lineinfile}, this issue can be easily fixed by adjusting the module's definition given in the module description language\footnote{We did not fix this simply because it is not clear if we should model the documented behavior (i.e., the actual behavior is a bug), the actual behavior (i.e., the behavior is intentional), or simply reject any program that triggers this behavior}.
Of the other 68 programs, 7 perform undesired additional actions that caused our tests to fail while the remaining 61 would work under certain assumptions not true in our testing environment; these issues are not unexpected given that as we have described formal queries are generally under-specified, these issues are entirely expected and as discussed in Section~\ref{sec:problem-statement} they are resolved by asking a user for input.
These 7 aforementioned programs are for problems 6 (creating a file if it does not exist) and 12 (rebooting the system if needed), which both involve conditionals like ``if $cond$ then do $foo$''.
Intuitively it seems such a query should also mean ``... otherwise do not do $foo$'' but since programs must be allowed to perform additional actions this is not the semantics of the formal query language.

Finally, the other 61 incorrect programs accepted by \system{} work under certain additional assumptions, which again is expected.
As an example, the five incorrect programs for problem 1 fail because they set the owner of the newly created directory to a user that did not exist.
The output of the unifier in these cases reports that they assume the user exists and they own the new directory, and a user could easily identify whether this matches their intent.
There is a similar issue with problem 17 where the two incorrect programs work if the ``service'' user does not already exist which a user can easily check.
The remaining issues involve the creation of files with certain contents, for example in problem 4 we are meant to copy a file on the remote machine while the generated code copies the file from the controller to the remote which is not rejected because the target files are ultimately created but it requires the alternate file sources to have certain contents; again these assumptions are identified by the unifier and a user could easily review them.

As discussed above, while \system{} is not always correct on whether a program is correct or not, its incorrect rejections are all addressable by adding support for additional features and the incorrect accepts are all expected as they are the result of the under-specified nature of the formal queries, the solution for them is to include a user to check whether the additional actions or assumptions they require match their intent.
These results prove that the techniques and ideas presented work and the failures are due to current limitations in our implementation.

\section{Discussion}\label{sec:discussion}

\subsection{Engineering Effort}\label{sec:engineering-effort}
\system{} has taken a significant amount of effort to develop and this could be a limiting factor to the adoption of our proposed approach.
\system{} was developed by one author in less than a year and while \system{} is still a research project and not fully featured, we believe that most of the additions to make it more full featured are relatively simple, such as writing specifications in the module language and handling the compilation of additional Ansible features.
Moreover, this investment of time and resources pales in comparison to the amount of time and resources required for the development of a LLM for code generation.

We also emphasize that this work is only a first step and
it has shown the strong potential of our approach.
We aim to reduce the necessary engineering effort in future work.
In particular, we believe there are opportunities to automate the construction of many components such as the knowledge base, parts of the formal query language, and components like module specification; one approach we are considering is to provide LLMs with documentation and other reference materials, asking them to generate the needed components, and then reviewing the results manually.

\subsection{Generality}\label{sec:generality}
While \system{} is specific to Ansible the ideas presented are relevant to other languages and the State Calculus and its symbolic interpreter may be useful as part of verifying other stateful languages.
In particular, we can already support a subset of bash programs and microcontroller programs, both of which are critical to correctness of many systems.
We use two examples to illustrate these capabilities, and leave it to future work to expand the support for each.

\paragraph{Bash Scripts}
Figure~\ref{fig:bash-example} shows a Bash script, a hand-translation of that script into the State Calculus, and a hand written specification in the State Calculus that could easily be derived from a formal query language for Bash.
Using these, and definitions of the Linux utilities used in this script, we have been able to use \system{} to verify that the hand-translation of the script matches the hand-written specification.
This example demonstrates a number of Bash features which are quite distinct from Ansible, including redirects, pipes, and even file descriptors.
One feature that we anticipate may be challenging to support are arbitrary loops; as described in Section~\ref{sec:interpreter} our handling of loops over symbolic values is pretty simple and does not support arbitrary while-loops or more complex forms of loops like found in languages like Bash.
We believe that even without this support, a significant and important subset of Bash can be supported; expanding to support these loops may be quite difficult as arbitrary loops are a challenge common challenge in verification.

\begin{figure}
    \centering
    \begin{subfigure}[b]{.59\textwidth}
        \centering
\begin{lstlisting}[language=Bash,basicstyle=\ttfamily\scriptsize]
tee >(wc -l >&2) < bigfile | gzip > bigfile.gz
\end{lstlisting}
        \caption{Example Bash script}
        \label{fig:bash-script}
\begin{lstlisting}[style=module]
let contents = fs('bigfile').contents;

env().stderr = ((env().stderr
    ^ string_of_int(count_lines(contents))
    ^ " ") ^ "\n";
fs('bigfile.gz').contents = gzipped(contents);
\end{lstlisting}
        \caption{Hand-written specification}
        \label{fig:bash-spec}
    \end{subfigure}
    \hfill
    \begin{subfigure}[b]{.4\textwidth}
        \centering
\begin{lstlisting}[style=module]
let pipe = new_pipe();
let file = tmp_file();
fd(0).kind = fd::file('bigfile');
fd(1).kind = fd::pipe(pipe);
tee {{ files: [file] }};
fd(0).kind = fd::file(file);
fd(1).kind = fd(2).kind;
wc {{ lines: true }};
fd(0).kind = fd::pipe(pipe);
fd(1).kind = fd::file('bigfile.gz');
gzip {{ }};
\end{lstlisting}
        \caption{Hand-translation of Bash script}
        \label{fig:bash-calculus}
    \end{subfigure}
    \caption{Example translation of a Bash script to the State Calculus. The script prints the number of lines in \texttt{bigfile} to stderr and gzips it to \texttt{bigfile.gz}.}
    \label{fig:bash-example}
\end{figure}

\paragraph{Arduino Programs}
Figure~\ref{fig:arduino-example} shows an Arduino\footnotemark{} program along with a hand-translation into the State Calculus and a hand-written specification that we are able to verify matches the hand-translation.
\footnotetext{Arduino is an open-source microcontroller, it is programmed in a C-like language~\cite{arduino-lang}.}
In this example we translate the main loop of the program which acts as a state machine and we simply verify that it acts correctly at each step.
In this setting, verification using \system{} may need to be paired with other verification techniques to ensure, for example, that expected invariants (such as assumptions about globals that arise from the State Calculus's unifier) are satisfied.

\begin{figure}
    \centering
    \begin{subfigure}[b]{.5\textwidth}
        \centering
\begin{lstlisting}[language=C,basicstyle=\ttfamily\scriptsize,numbers=left]
const int BUTTON = 2;
const int LED = 8;
int buttonState;
int ledState;
void setup() {
    pinMode(LED, OUTPUT);
    pinMode(BUTTON, INPUT);
    ledState = LOW;
    digitalWrite(LED, ledState);
    buttonState = digitalRead(BUTTON);
}
void loop() {
    int oldState = buttonState;
    buttonState = digitalRead(BUTTON);

    if  (oldState==LOW&&buttonState==HIGH) {
        ledState = !ledState;
        digitalWrite(LED, ledState);
    }
}
\end{lstlisting}
        \caption{Example Arduino program}
        \label{fig:arduino-script}
    \end{subfigure}
    \hfill
    \begin{subfigure}[b]{.49\textwidth}
        \centering
\begin{lstlisting}[style=module,escapechar=@]
let time = any_time();
let oldState = globals().buttonState;
globals().buttonState = digitalRead(2);
if !oldState && globals().buttonState {
    globals().ledState = ! globals().ledState;
    digitalWrite(8, globals().ledState);
}
        \end{lstlisting}
        \caption{Hand-translation of Arduino program}
        \label{fig:arduino-calculus}
        \begin{lstlisting}[style=module,escapechar=@]
let cur_time = any_time();
let prev_time = cur_time - 1;
if !input(prev_time).pin(2).pinValue
&& input(cur_time).pin(2).pinValue {
    output(cur_time).pin(8).pinValue =
        ! output(prev_time).pin(8).pinValue;
}
\end{lstlisting}
        \caption{Hand-written specification}
        \label{fig:arduino-spec}
    \end{subfigure}
    \caption{Example translation of an Arduino program to the State Calculus. The program toggles an LED each time a button is clicked.}
    \label{fig:arduino-example}
\end{figure}

We highlight that in both examples, like with Ansible, the translation from the source language to the Calculus is straightforward, the main source of complexity comes from the language specification, for Bash this includes the utilities and for Arduino the built-in functions.
While both of these examples rely on hand-translation of the source programs and there is still significant work to be done on determining the best way to represent the state for these languages, these examples strongly suggest to us that the State Calculus can be used to verify the behavior of Bash and Arduino programs and likely programs in other similar languages as well.

\section{Related Work}\label{sec:related-work}
\noindent\paragraph{Controlled Natural Languages}
There has been significant research on Controlled Natural Languages (CNLs)~\cite{ace.case99,peng.desa02,cnl.cl14,conceptualAuthoring.cl07} which have a syntax based on a restricted form of natural language.
CNLs have been used in various applications, from database query languages~\cite{conceptualAuthoring.cl07} to first-order logic specifications~\cite{ace.case99,peng.desa02}.
The formal query languages we propose in this work are similar to CNLs, which bolsters our belief that formal queries can be easily understood by users.
To our knowledge, CNLs have not been used before to formalize user intent for LLM code-generation.

\noindent\paragraph{Identifying Incorrect LLM Generated Code}
Some prior work has used test cases to check LLM generated code~\cite{codex.arxiv21}, which increases the likelihood that the generated code is correct and with a sufficient test suite could offer strong assurances of correctness but this requires the user to craft good tests; which is so hard that errors were found in the ground-truth solutions to the well known \textsc{HumanEval} benchmark suite~\cite{evalPlus.neurips23} despite their tests.

\noindent\paragraph{Verified LLM Generated Code}
LLM code generation has been explored for generating proofs in proof assistants such as Rocq and Isabelle~\cite{baldur.fse23} because the LLM generated proof can be checked for correctness by the theorem prover, but this approach does not generalize to general code where oracles for verifying correctness do not exist.
This work also relies on a user specified formal theorem statement,
though auto-formalization has been used to convert natural language theorem statements into formal ones~\cite{autoformalization.neurips22}, but the generated theorem statements are in a lower-level language than our formal query language making them more difficult for users to check.

\noindent\paragraph{Natural Language Summaries}
Showing the user a natural language-like description to check has been proposed before but prior work has explored providing a description of a generated program~\cite{photoscout.chi24,groundedAbstraction.chi23}.
While this can help a user determine if the generated code matches their intent we believe our approach of having the user verify a high-level specification is better since it helps avoid issues where a code summary may seem reasonable but contain subtle errors.

\noindent\paragraph{LLM Generation of Formal Specifications}
The use of LLMs to formalize a user's intent has been explored in other contexts such as generating post-conditions~\cite{nl2postcond.fse24} and generating Isabelle theorem statements~\cite{autoformalization.neurips22}.
These works show that LLMs can generate reasonable formalizations which can then be used to check generated code or proofs.
However, we believe that the formalizations chosen by these prior works are not suited for the user to check correctness of as these are significantly lower-level than the the natural language description.

\noindent\paragraph{Program Synthesis}
Finally, our work is an attempt to merge LLM-based code generation with ideas and techniques from Program Synthesis.
In particular, in our vision for a complete code-generation system (Figure~\ref{fig:system-diagram}) the loop between the Code Generator and Verifier bears significant resemblance to the feedback loop found in Counterexample Guided Inductive Synthesis, and similar techniques.
Additionally, the user's input in the Assumptions Check phase bears resemblance to interactive synthesis~\cite{iterative-synth.lntcs17} where the user may iteratively refine the specification until the generated program matches their full intent.
There has also been some prior work that makes use of both LLM code generation and program synthesis, such as Photoscout~\cite{photoscout.chi24} which uses an LLM to generate programs and then a semantic analysis identifies undefined components and replaces them with holes which are filled using an enumerative synthesis approach.
This technique could be relevant to Ansible, as our formal queries could provide a basis to perform such synthesis.

\section{Conclusion}\label{sec:conclusion}
In this work we have proposed the use of \emph{formal query languages}, which formalize a user's intent, to check LLM generated code for correctness.
We present design constraints for such languages, to ensure a user can check that a query matches their intent, and present our implementation, \system{}, for the Ansible IT automation language and show that we can use our verifier to effectively identify correct and incorrect LLM generated code.
While the verifier is imperfect, we analyze the errors and find that the errors are all results of limitations in our testing setup or in the implementation that could be easily addressed.
We believe this work can lay groundwork for building systems which, from a natural language description, are able to generate code that is guaranteed to match the user's intent, which is essential to allow AI Code Assistants to be adopted in critical systems.

\begin{acks}
    This work is supported by the IBM-ILLINOIS Discovery Accelerator Institute (IIDAI). Any opinions, findings, and conclusions expressed in this material are those of the authors and do not necessarily reflect the views of IBM.
    
    This research used the Delta advanced computing and data resource which is supported by the National Science Foundation (award OAC 2005572) and the State of Illinois. Delta is a joint effort of the University of Illinois Urbana-Champaign and its National Center for Supercomputing Applications.
    This use was through allocation CIS250276 from the Advanced Cyberinfrastructure Coordination Ecosystem: Services \& Support (ACCESS) program, which is supported by U.S. National Science Foundation grants \#2138259, \#2138286, \#2138307, \#2137603, and \#2138296.
\end{acks}

\bibliographystyle{ACM-Reference-Format}
\bibliography{references}

\appendix

\section{Benchmark Suite Natural Language Descriptions and Formal Queries}\label{app:benchmark-suite}
\begin{tabularx}{\textwidth}{|c|X|X|} \hline
    \textbf{\#} &\textbf{Natural Language Desciption} &\textbf{Formal Language Query} \\ \hline
    1 &Create directory /srv/www
      &create directory at /srv/www \\ \hline
    2 &Delete the directory /home/mydata/web
      &delete directory at /home/mydata/web \\ \hline
    3 &Delete the contents of the /home/mydata/web directory
      &delete files in /home/mydata/web \\ \hline
    4 &Copy the file /scratch/file.txt to /home/user/data.txt
      &copy file from "/scratch/file.txt" to "/home/user/data.txt" \\ \hline
    5 &Move the file at /scratch/file.txt to /home/user/data.txt
      &move file from "/scratch/file.txt" to "/home/user/data.txt" \\ \hline
    6 &Create the file /etc/file.txt with the contents ``beginning'' if it does not already exist
      &if file "/etc/file.txt" not exists then create file at "/etc/file.txt" with content="beginning" \\ \hline
    7 &Copy all .war files from /backup on the controller into /tmp on the remote
      &copy files with glob="*.war" from controller /backup to remote /tmp \\ \hline
    8 &Copy all .war files from /tmp on the remote machine into /backup on the controller
      &copy files with glob="*.war" from remote /tmp to controller /backup \\ \hline
    9 &Clone the release1.8.4 branch from the ansible/ansible github repo to /ansible using https
      &clone github repository with name=ansible/ansible, branch="release1.8.4" into /ansible via https \\ \hline
    10 &Permanently set the environment variable LC\_ALL to ``C'' system-wide
      &set environment variable LC\_ALL to "C" \\ \hline
    11 &Add the line `module load gcc' to the end of foo's bashrc
      &write "module load gcc" to bashrc file for user = foo at position=end \\ \hline
    12 &For Debian, restart the system if it is needed
      &if os is Debian and reboot required then reboot \\ \hline
    13 &Set permissions for the files and directories under /srv/path so that all users can list and read all files but only the owner can write
      &set file permissions in /srv/path to read=all, write=owner, list directory=all \\ \hline
    14 &Create a Python 3.10 virtual environment at /user/home/venv and install numpy in it
      &create virtual environment with python="3.10" in /user/home/venv; install numpy in virtual environment at /user/home/venv \\ \hline
    15 &For Debian and RedHat based systems, install numpy
      &if os is Debian based or os is RedHat based then install numpy \\ \hline
    16 &Install zsh, set it as the user `dev''s default shell, and ensure its configuration file exists
      &install zsh; set default shell for user="dev" to zsh; if zsh configuration file for user="dev" not exists then create zsh configuration file for user="dev" \\ \hline
    17 &Disable the password for the `service' user
      &disable password for user=service \\ \hline
    18 &For RedHat and Ubuntu, install the lastest version of postfix, backup its default configuration file, and download the configuration file from http://example.com/cfg.ubuntu for Ubuntu and http://example.com/cfg.redhat for RedHat
      &if os is RedHat or os is Ubuntu then install postfix with version=latest; copy postfix configuration file to ?backup. if os is Ubuntu then download postfix configuration file from "http://example.com/cfg.ubuntu". if os is "RedHat" then download postfix configuration file from "http://example.com/cfg.redhat" \\ \hline
    19 &For RedHat, install apache server, ensure the home page exists, and create a `webdev' user
      &if os is RedHat then install apache server; create user with name=webdev; if apache server html home page file not exists then create apache server html home page file \\ \hline
    20 &Enable passwordless sudo for a `wheel' group and create an `ansible' user in that group
      &enable passwordless sudo for group=wheel; create user with name=ansible in supplemental groups=wheel \\ \hline
    21 &For Debian and RedHat, install and start an ssh client and server and generated an ssh key for `fizbaz' at /.ssh/fizbaz\_rsa
      &if os is Debian or os is RedHat then install ssh client; install ssh server; start ssh server service; create ssh key for user="fizbaz", name="fizbaz\_rsa" \\ \hline
\end{tabularx}

\section{LLM Code-Generation Prompt}\label{app:prompt}

\begin{tcolorbox}
You are an expert in Ansible. The user will give you a task description and ask you to generate an Ansible playbook to complete the given task. You only need to output the content of the playbook. DO NOT use any shell commands (ansible.builtin.shell, ansible.builtin.command, etc.) in the playbook.

Task: \{ task \}

Answer: \textasciigrave\textasciigrave\textasciigrave{}yaml
\end{tcolorbox}

\end{document}